\documentstyle[twocolumn,prc,aps,epsf]{revtex}

\preprint{NUC-MINN-02/8-T}
\newcommand{\be}{\begin{equation}}
\newcommand{\ee}{\end{equation}}

\newcommand{\ba}{\begin{eqnarray}}
\newcommand{\ea}{\end{eqnarray}}

\draft
\begin{document}

\title{Optically opaque color-flavor locked phase inside compact stars} 

\author{Igor A.~Shovkovy$^{1,2,}$\thanks{On leave from 
        Bogolyubov Institute for Theoretical Physics,
        03143 Kiev, Ukraine.}
and Paul J. Ellis$^{2}$}

\address{$^{1}$Institut f\"ur Theoretische Physik,
Johann Wolfgang Goethe--Universit\"at,
60054 Frankfurt/Main, Germany}

\address{$^{2}$School of Physics and Astronomy, University of Minnesota,
        Minneapolis, Minnesota 55455}

\date{\today}

\maketitle

\begin{abstract} 

The contribution of thermally excited electron-positron pairs to the bulk
properties of the color-flavor locked quark phase inside compact stars is
examined. The presence of these pairs causes the photon mean free path to
be much smaller than a typical core radius ($R_{0} \simeq 1$ km) for all
temperatures above $25$ keV so that the photon contribution to the thermal
conductivity is much smaller than that of the Nambu-Goldstone bosons.  We
also find that the electrons and positrons dominate the electrical
conductivity, while their contributions to the total thermal energy is
negligible.

\end{abstract}



While at present there is no unequivocal evidence that free quarks exist
in compact stars, it is important to examine the role they might play
since the observational situation may well change in the future. If quarks
are present in the center of compact stars, it is most likely that they
will be found in the color-flavor locked (CFL) phase \cite{ARW}. There
exists a rather detailed understanding of the basic properties of CFL
quark matter \cite{ARW,CasGat,SonSt,ShoWij,Risch,other,rev}. This brief
report augments our previous discussion \cite{cool-star} of the thermal
properties of CFL quark matter by addressing the role of thermally excited
electron-positron pairs which, hitherto, has been overlooked.

We start by mentioning that it is now commonly accepted that the CFL phase
is electrically neutral \cite{neutral}. This would suggest that the
chemical potential related to the electric charge, $\mu_e$, is zero;
implying that at finite temperature the (nonvanishing)  densities of
electrons and positrons are equal. In fact, Lorentz invariance is broken in
the CFL phase and the positively and negatively charged kaons differ in
mass, as do pions \cite{bs}.  Therefore, at finite temperature, the
densities of the positively and negatively charged species are not exactly
the same; and this must be balanced by differing electron and positron
densities in order to maintain charge neutrality. Thus $\mu_e$ is
nonzero. However, at temperatures below $5$ MeV, $\mu_e$ drops rapidly to
zero \cite{rst}, and it is completely negligible at temperatures of $1$ MeV
or less, which are our principal interest here. It is therefore sufficient
to set $\mu_e=0$ in assessing the impact of electron-positron pairs on the
physical properties of CFL quark matter.

The first issue to address is the photon mean free path. Since photons
scatter quite efficiently from charged leptons, even small numbers of
electrons and positrons could substantially reduce the transparency of CFL
quark matter in the core of a compact star.  Now the photon mean free path
can be rather well approximated by the simple expression
\be
\ell_{\gamma} \simeq \frac{1}{2 n_{e}(T) \sigma_{\rm T}}\;,
\ee
where $n_{e}(T)$ is the equilibrium density of electrons in a plasma at
temperature $T$, the factor 2 takes into account the equal density of
positrons, and
\be
\sigma_{\rm T} = \frac{8\pi}{3} \frac{\alpha^{2}}{m_{e}^{2}}
\approx 66.54 \mbox{~fm}^{2}
\ee
is the well-known expression for the Thomson cross section in terms of the
fine structure constant $\alpha$ and the electron mass $m_e$. This
expression is the limiting case of the more complicated Compton cross
section for low photon energies. Since this limit works rather well for
$\omega_{\gamma} \ll m_{e}$ and $m_{e} \simeq 0.5$ MeV, this is sufficient
for the purposes of this paper.

\begin{figure}
\begin{center}
\epsfxsize=8.0cm
\epsffile[88 20 595 285]{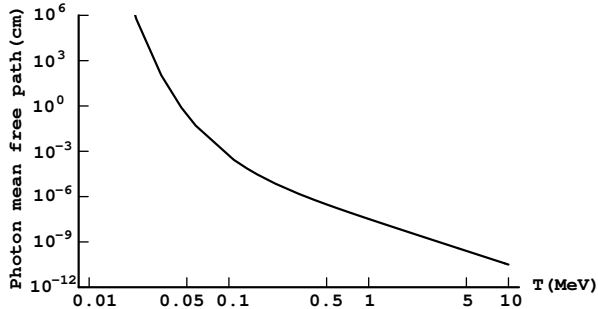}
\caption{Photon mean free path as a function of temperature 
in the color-flavor locked phase.}
\label{fig-mfp}
\end{center}
\end{figure}

The average equilibrium electron or positron density in a finite
temperature neutral plasma \cite{thermo-asympt} is
\be
n_e = \frac{m_{e}^2 T}{\pi^{2}} \sum_{k=1}^{\infty}
\frac{(-1)^{k+1}}{k} K_{2}\!\left(\frac{m_{e}k}{T}\right),
\label{n_e}
\ee
where $K_2$ is a modified Bessel function. By making use of this result we
find the temperature dependence of the photon mean free path in the CFL
phase shown in Fig.~\ref{fig-mfp}. We see that $\ell_{\gamma} \alt 220$ m
for $T \agt 25$ keV. Since the radius of the CFL core ($R_0$) is of order
1 km, the photon mean free path is short for temperatures above $25$ keV,
so that the quark core of a compact star is opaque to light. Conversely,
transparency can be considered to set in when the mean free path exceeds
$1$ km, which occurs for temperatures below $23.4$ keV.

Photons together with massless Nambu-Goldstone (NG) bosons 
$\phi$, which arise from the
breaking of the global baryon number symmetry, dominate the thermal
conductivity. The estimates above indicate that for $T \alt 25$ keV, these
components are equally important, as discussed in our previous study
\cite{cool-star}. This is no longer true at higher temperatures since the
photon contribution becomes negligible. Thus the thermal conductivity can
be approximated,
\be
\kappa_{\rm CFL}\simeq\kappa_{\phi}
\simeq \frac{2\pi^{2}}{15} T^{3} R_{0}
\quad \mbox{for} \quad T \agt 25\mbox{~keV},
\label{kappa-CFL}
\ee
where for the velocity we used $v_{\phi}=1/\sqrt{3}$. Numerically, this 
leads to the following estimate
\be
\kappa_{\rm CFL} \simeq (7.2 \times 10^{31})~ 
T_{\rm MeV}^{3} R_{0,{\rm km}}
\mbox{~erg~cm}^{-1} \mbox{~sec}^{-1} \mbox{~K}^{-1},
\label{kappa-CFL-num}
\ee
where the notation indicates that $T$ is measured in units of MeV and
$R_0$ is measured in units of km. This differs from our previous result
\cite{cool-star} by a factor $3/5$, which does not change the qualitative
conclusion that the thermal conductivity of CFL matter is extremely high.

\begin{figure}
\begin{center}
\epsfxsize=8.0cm
\epsffile[88 20 580 280]{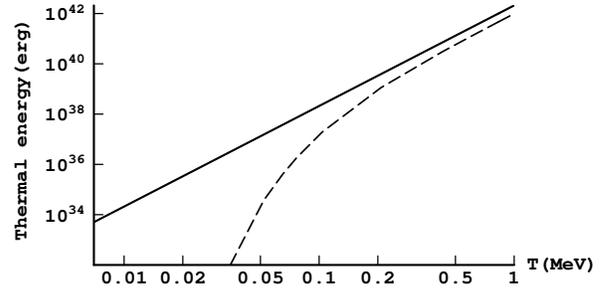}
\caption{The contributions of massless states (solid line) and 
charged leptons (dashed line) to the thermal energy of a CFL quark 
core with radius $R_{0}=1$ km.}
\label{fig-th-en}
\end{center}
\end{figure}

We also need to evaluate the electron-positron contribution to the thermal
energy of the CFL core since this is relevant to the cooling. It is 
straightforward to obtain the expression (see Ref.~\cite{thermo-asympt})
\ba
E_{e}(T) &=& \frac{8T \left(R_{0} m_{e}\right)^{3}}{3\pi}
\sum_{k=1}^{\infty}\frac{(-1)^{k+1}}{k} \nonumber \\
&\times& \left[K_{1}\!\left(\frac{m_{e}k}{T}\right)
+\frac{3T}{m_{e}k} K_{2}\!\left(\frac{m_{e}k}{T}\right)\right].
\label{E_e}
\ea
In Fig.~\ref{fig-th-en} we indicate this contribution to the thermal
energy by a dashed line, while our previously calculated thermal energy
due to the massless states, $E_{\phi,\gamma}$, is indicated by the solid
line (the radius of the CFL quark core, $R_0$, is taken to be $1$ km).
The comparison shows that the charged lepton contribution is very small,
$E_{e} \ll E_{\phi,\gamma}$, for all temperatures below $0.1$ MeV.  
Moreover, even at higher temperatures, $E_{e} \alt E_{\phi,\gamma}$.  
Thus, the presence of thermally excited electron-positron pairs does not
qualitatively affect the thermal energy of CFL matter. This finding,
together with the previous discussion of the thermal conductivity, means
that the cooling mechanism described in Ref.~\cite{cool-star} remains
unchanged.

A separate issue is the electrical conductivity of CFL matter. Since
electrons and positron are charged particles, they should give a
nonvanishing contribution to this transport coefficient.  In principle,
charged pseudo-NG bosons will also contribute, as mentioned in
Ref.~\cite{cool-star}. However, since even the lightest $K^{+}$ boson has
an estimated mass that is an order of magnitude larger than that of the
electron (e.g., see Refs.~\cite{SonSt,kaon-mass}), these
contributions will be strongly suppressed and electrons and positrons 
will dominate the electrical conductivity.

In order to estimate the electrical conductivity of the thermally excited
electron-positron plasma in the CFL phase, we need to know the temperature
dependence of the electron mean free path. The mean free path is finite
due to the scattering of electrons by positrons and other electrons, as
well as annihilation, braking radiation (bremsstrahlung), and Compton
scattering. These processes have recently been studied in the
ultrarelativistic limit in order to determine the kinetic properties of
gauge theories \cite{AMY}. However, in the non-relativistic limit, the
annihilation, braking radiation, and Compton scattering contributions are
expected to be negligible, and the corresponding expression for the
electron mean free path is known in the so-called leading logarithm
approximation. This is sufficient if we restrict ourselves to temperatures
$T\alt m_{e}$, in which case the electron mean free path \cite{LifPit} is
\be 
\ell_{e}\sim \frac{T^{2}}{4\pi \alpha^{2} n_{e} L_{e} }, 
\label{el-mfp}
\ee 
where $n_{e}$ is the number density of positrons which, according to our
assumption, is the same as that of electrons given in Eq.~(\ref{n_e}). The
so-called Coulomb logarithm $L_{e}$ appears as a result of the long
range nature of the Coulomb interaction. Its value is determined by the
Debye screening mass $m_{D}$ inside the plasma \cite{LifPit}:
\be 
L_{e} \simeq \ln\frac{T}{m_{D}\mbox{max}(\alpha, \bar{v}_{e})}, 
\quad \mbox{where} \quad m_{D}^{2}=\frac{8\pi \alpha n_{e}}{T},
\label{L-e} 
\ee
and $\bar{v}_{e}\simeq \sqrt{T/m_{e}}$ is the average thermal velocity of
the electrons.

Now, to estimate the electrical conductivity of the electron-positron
plasma existing inside the CFL quark core of the star, we use the 
following classical expression:
\be 
\sigma_{e} \simeq \frac{8\pi\alpha n_{e} \ell_{e} }{ m_{e} \bar{v}_{e}},
\label{el-cond} 
\ee
relating the conductivity (in Heaviside-Lorentz units) to the electron
mean free path $\ell_{e}$. If $\ell_{e}$ is less than the typical size 
of the CFL quark phase, $R_{0}\simeq 1$ km, which is the case for
temperatures higher than about $17$ keV, this expression for the
electrical conductivity reads
\be 
\sigma_{e} \sim \frac{2T^{3/2}}{\alpha \sqrt{m_{e}} L_{e} }. 
\ee
It should be noted that this expression depends on the density of
electrons only through the Coulomb logarithm $L_e$. If this expression
and the ultrarelativistic result \cite{AMY} are extrapolated to $T\sim
m_e$, it is satisfactory that they are of similar magnitude.

At lower temperatures ($T \alt 17$ keV), on the other hand, the 
electrical conductivity in Eq. (\ref{el-cond}) becomes
\be
\sigma_{e} \simeq \frac{2^{5/2} \alpha m_{e} T R_{0}} { \sqrt{\pi} }
\exp\left(-\frac{m_{e}}{T}\right).
\label{el-cond-R0}
\ee
This makes it clear that the contribution from, say, a positively charged
kaon will be enormously suppressed due to a factor $\exp
\left(-m_{K^{+}}/T\right)$.

Note that the contribution of the electrons and positrons themselves to
the thermal conductivity is always small compared to the contribution of
NG bosons in Eq.~(\ref{kappa-CFL}). At temperatures higher than about $17$
keV, it is small because the electron mean free path is small, while at
lower temperatures $\ell_{e}$ is restricted by the size of the core, and
the specific heat decreases exponentially with decreasing temperature.

In conclusion, we have shown that the presence of a thermally excited
electron-positron plasma inside the neutral CFL core of a compact star has
some interesting consequences. One of the most interesting, and somewhat
unexpected, consequences is that CFL quark matter is optically opaque at
temperatures higher than about $25$ keV. This suppresses the photon
contribution to the thermal conductivity, although at lower temperatures
it is comparable to that arising from the massless NG boson, $\phi$.
Nevertheless, the thermal conductivity of CFL matter is very large in all
regimes, so that the thermal energy from the CFL core of a compact star is
efficiently conducted away to the outer nuclear layer and the core remains
nearly isothermal \cite{cool-star}. If bare CFL quark stars exist in the
Universe, the fact that they only become transparent to photons when they
have cooled to rather low temperatures might be of some observational
importance. A similar interplay between NG boson and photon contributions
should also appear in other transport coefficients, such as the shear
viscosity.

The electrical conductivity of the CFL phase will be largely determined by
the electron-positron plasma since contributions from other charged
particles, such as kaons, are strongly suppressed. This may turn out to be
important in studying the dynamo mechanism \cite{dynamo} of magnetic field
generation in compact stars with quark cores.\\

We thank Thomas Sch\"{a}fer 
for comments on the first version of the paper. 
This work was supported by the U.S. Department of
Energy Grant No.~DE-FG02-87ER40328. The work of I.A.S. was partially
supported by the Gesellschaft f\"{u}r Schwerionenforschung (GSI) and by
the Bundesministerium f\"{u}r Bildung und Forschung (BMBF).

\end{document}